\documentclass[prd,preprint,nofootinbib]{revtex4}
\usepackage{amssymb}
\usepackage{epsfig}

\begin{document}
\title{ 
Neutralino dark matter in brane world cosmology
}
\author{Takeshi Nihei}
 \email{nihei@phys.cst.nihon-u.ac.jp}
 \affiliation{
  Department of Physics, College of Science and Technology, Nihon University,
  1-8-14, Kanda-Surugadai, Chiyoda-ku, Tokyo 101-8308, Japan
 }

\author{Nobuchika Okada}
 \email{okadan@post.kek.jp}
 \affiliation{
  Theory Division, KEK, Oho 1-1, Tsukuba, Ibaraki 305-0801, Japan
 }

\author{Osamu Seto \footnote{
 Present address: Department of Physics and Astronomy, University of Sussex, 
 Brighton BN1 9QJ, United Kingdom
 }}
 \email{O.Seto@sussex.ac.uk}
 \affiliation{
 Institute of Physics, National Chiao Tung University, 
 Hsinchu, Taiwan 300, Republic of China
}


\begin{abstract}
The thermal relic density of neutralino dark matter
 in the brane world cosmology is studied. 
Since the expansion law at a high energy regime 
 in the brane world cosmology 
 is modified from the one in the standard cosmology, 
 the resultant relic density can be altered. 
It has been found that, 
 if the five dimensional Planck mass $M_5$ is lower than $10^4$ TeV, 
 the brane world cosmological effect is significant 
 at the decoupling time, and the resultant relic density is enhanced. 
We calculate the neutralino relic density 
 in the constrained minimal supersymmetric standard model 
 and 
 show that the allowed region is dramatically modified 
 from the one in the standard cosmology 
 and eventually disappears as $M_5$ is decreasing. 
We also find a new lower bound on $M_5 \gtrsim 600$ TeV 
 based on the neutralino dark matter hypothesis, 
 namely, the lower bound in order 
 for the allowed region of neutralino dark matter to exist. 
\end{abstract}

\preprint{KEK-TH-982 }

\vspace*{1cm}

\maketitle


\section{INTRODUCTION}

Recent cosmological observations,
 especially the Wilkinson Microwave Anisotropy Probe (WMAP) 
 satellite \cite{WMAP}, 
 have established the $\Lambda$CDM cosmological model
 with great accuracy, and 
 the relic abundance of the cold dark matter is estimated as
 (in 2 $\sigma$ range) 
\begin{equation}
\Omega_{CDM} h^2 = 0.1126^{+0.0161}_{-0.0181} .
\end{equation}
To clarify the identity of a particle as cold dark matter 
 is still an open prime problem in cosmology and particle physics. 
The lightest supersymmetric particle (LSP) is suitable 
 for cold dark matter, 
 because they are stable owing to the conservation of R-parity.
In the minimal supersymmetric standard model (MSSM), 
 the lightest neutralino is typically the LSP 
 and the promising candidate for cold dark mater. 
In light of the WMAP data, 
 the parameter space in the constrained MSSM (CMSSM) 
 which allows the neutralino relic density 
 suitable for cold dark matter 
 has been recently re-analyzed \cite{CDM}. 
It has been shown that the resultant allowed region 
 is dramatically reduced due to the great accuracy 
 of the WMAP data. 

Note that the thermal relic density of the matter 
 depends on the underlying cosmological model 
 as well as its annihilation cross section. 
If a nonstandard cosmological model is taken into account, 
 the resultant relic density of the dark matter can be altered 
 from the one in the standard cosmology. 
A brane world cosmological model 
 which has been intensively investigated \cite{braneworld} 
 is a well-known example as such a nonstandard cosmological model. 
The model is a cosmological version 
 of the so-called ``RS II'' model 
 first proposed by Randall and Sundrum \cite{RS}, 
 where our four-dimensional universe is realized 
 on the ``3-brane'' located at the ultra-violet boundary 
 of a five dimensional Anti-de Sitter spacetime. 
In this setup, the Friedmann equation for a spatially flat spacetime
 is found to be
\begin{equation}
H^2 = \frac{8\pi G}{3}\rho\left(1+\frac{\rho}{\rho_0}\right),
\label{BraneFriedmannEq}
\end{equation}
where
\begin{eqnarray}
\rho_0 = 96 \pi G M_5^6,
\label{def:rho_0}
\end{eqnarray}
 $H$ is the Hubble parameter, $\rho$ is the energy density of matters, 
 $G$ is Newton's gravitational constant 
 with $M_5$ being the five-dimensional Planck mass, 
 and the four-dimensional cosmological constant has been 
 tuned to be almost zero. 
Here we have omitted the term so-called ``dark radiation'', 
 since it is severely constrained by 
 Big Bang Nucleosynthesis (BBN) \cite{ichiki}. 
The coefficient $\rho_0$ is also constrained  by the BBN, 
 which is roughly given by $\rho_0^{1/4} \gtrsim 1$ MeV 
 (or, equivalently, $M_5 \gtrsim 8.8$ TeV) \cite{braneworld}.
This is a model-independent cosmological constraint. 
On the other hand, 
 as discussed in the original paper by Randall and Sundrum \cite{RS}, 
 the precision measurements of the gravitational law 
 in sub-millimeter range lead to more stringent constraint 
 $\rho_0^{1/4} \gtrsim 1.3$ TeV 
 (or $M_5 \gtrsim 1.1 \times 10^8$ GeV)  
 through the vanishing cosmological constant condition. 
However note that this constraint, in general, 
 is quite model dependent. 
In fact, if we consider an extension of the model 
 so as to introduce a bulk scalar field, 
 the constraint can be moderated 
 because of the change of 
 the vanishing cosmological constant condition \cite{maedawands}. 
Hence, we care about only the BBN constraint 
 on $\rho_0$ in this paper. 

Note that the $\rho^2$ term in Eq.~(\ref{BraneFriedmannEq}) 
 is a new ingredient in the brane world cosmology. 
Since at a high energy regime this term dominates and 
 the universe obeys a nonstandard expansion law, 
 some results previously obtained in the standard cosmology 
 can be altered. 
In fact, some interesting consequences 
 in the brane world cosmology 
 have been recently pointed out \cite{OS,OS2}. 
Especially, 
 the thermal relic abundance of dark matter 
 can be considerably enhanced 
 compared to that in the standard cosmology \cite{OS}. 

In this paper, 
 we investigate the brane world cosmological effect 
 for the relic density of the neutralino dark matter in detail 
 by numerical analysis. 
We will show that 
 the allowed region for the neutralino dark matter in the CMSSM 
 is dramatically modified in the brane world cosmology. 
In the next section, 
 we give a brief review of Ref.~\cite{OS}. 
In sec.~III, we present our numerical results 
 for the neutralino relic density in the brane world cosmology. 
We will find some interesting consequences. 
Sec.~IV is devoted to conclusions.

\section{Enhancement of relic density in brane world cosmology}

In this section, we give a brief review 
 on the relic density of dark matter 
 in the brane world cosmology 
 with a low five-dimensional Planck mass \cite{OS}. 

In the context of the brane world cosmology, 
 we estimate the thermal relic density of 
 a dark matter particle 
 by solving the Boltzmann equation
\begin{equation}
\frac{d n}{d t}+3Hn = -\langle\sigma v\rangle(n^2-n_{EQ}^2),
\label{n;Boltzmann}
\end{equation}
with the modified Friedmann equation Eq.~(\ref{BraneFriedmannEq}),
where $n$ is the actual number density of the dark matter particle,
 $n_{EQ}$ is the equilibrium number density,
 $\langle\sigma v\rangle$ is the thermal averaged product
 of the annihilation cross section $\sigma$ and the relative velocity $v$.
It is useful to rewrite Eq.~(\ref{n;Boltzmann}) 
 into the form,  
\begin{eqnarray}
\frac{d Y}{d x}
&=& -\frac{s}{xH}\langle\sigma v\rangle(Y^2-Y_{EQ}^2) \nonumber\\
&=&- \lambda\frac{x^{-2}}{\sqrt{1+ \left(\frac{x_t}{x} \right)^4}}
\langle\sigma v\rangle(Y^2-Y_{EQ}^2) ,
\label{Y;Boltzmann}
\end{eqnarray}
in terms of the number density to entropy ratio $Y = n/s$ and $x = m/T$, 
 where $m$ is a dark matter particle mass, 
 $\lambda = 0.26 (g_{*S}/g_*^{1/2}) M_{P} m$, 
 $M_P \simeq 1.2 \times 10^{19}$GeV is the Planck mass, 
 and $x_t$ is defined as 
\begin{equation}
 x_t^4 \equiv \left. \frac{\rho}{\rho_0}\right|_{T=m}. 
\label{def:xt}
\end{equation}

At the era $x \ll x_t$ the $\rho^2$ term dominates 
 in Eq.~(\ref{BraneFriedmannEq}), 
 while the $\rho^2$ term becomes negligible after $x \gg x_t$ 
 and the expansion law in the standard cosmology is realized. 
Hereafter, we call the temperature defined 
 as $T_t = m x_t^{-1}$ (or $x_t$ itself) ``transition temperature''
 at which the expansion law of the early universe changes 
 from the non-standard one to the standard one. 
Since we are interested in the effect of the $\rho^2$ term 
 for the dark matter relic density, 
 we consider the case that the decoupling temperature 
 of the dark matter ($T_d$) is higher than the transition temperature, 
 namely, $x_t \geq x_d=m/T_d$. 
Using Eqs.~(\ref{def:rho_0}) and (\ref{def:xt}), 
 this condition leads to 
\begin{eqnarray}
M_5 & \leq & \left( \frac{\pi^2 g_*}{30} m^4 \frac{1}{96\pi G}
 x_d^{-4} \right)^{1/6}  
\\
&\simeq& 4.6 \times 10^3 \,\textrm{TeV}\, 
\left(\frac{m}{100\, \textrm{GeV}}\right)^{2/3} 
\left(\frac{20}{x_d}\right)^{2/3} .
\label{critM5:100}
\end{eqnarray} 
Here we have normalized the decoupling temperature by 
 its typical value in our case estimated 
 as $x_d \equiv m/T_d \simeq 20$, 
 which is the same scale as in the standard cosmology \cite{Kolb}, 
 as is shown in the appendix. 
For $M_5 \lesssim 10^3$ TeV, 
 we can expect significant brane world cosmological effects. 
This, indeed, will be confirmed by numerical calculations 
 in the next section (see Fig.~3). 

It is easy to numerically solve 
 the Boltzmann equation Eq.~(\ref{Y;Boltzmann}) 
 with a given $\langle \sigma v \rangle$ and $x_t$. 
Using appropriate approximations, 
 we can drive analytic formulas 
 for the relic number density of dark matter \cite{OS}. 
When we parameterize $\langle \sigma v \rangle$ 
 as $\langle \sigma v \rangle = \sigma_n x^{-n}$ 
 with fixed $n=0,1,\cdots$, for simplicity, 
 we can obtain simple formulas 
 for the resultant relic densities 
 such that 
\begin{eqnarray}
 Y(x \rightarrow \infty) 
 &\simeq& 0.54 \frac{x_t}{\lambda \sigma_0} 
 \qquad \textrm{for} \quad n=0, \nonumber \\
 && \frac{x_t^2}{\lambda \sigma_1 \ln x_t} \qquad \textrm{for} \quad n=1, 
\label{Ybrane}
\end{eqnarray}
in the limit $x_d \ll x_t$, 
 where $x_d$ is the decoupling temperature. 
Note that the results are characterized 
 by the transition temperature 
 rather than the decoupling temperature. 
It is interesting to compare these results 
 to that in the standard cosmology. 
Using the well-known approximate formulas 
 in the standard cosmology \cite{Kolb},  
\begin{eqnarray}
Y(x \rightarrow \infty) 
 &\simeq & \frac{x_d}{\lambda \sigma_0} 
 \qquad \textrm{for} \quad n=0, \nonumber \\ 
 && \frac{2 x_d^2}{\lambda \sigma_1} 
 \qquad \textrm{for} \quad n=1,  
\label{Ystandard}
\end{eqnarray} 
we obtain the ratio of the relic energy density 
 of dark matter in the brane world cosmology ($\Omega_{(b)}$) 
 to the one in the standard cosmology ($\Omega_{(s)}$) 
 such that 
\begin{eqnarray}
\frac{\Omega_{(b)}}{\Omega_{(s)}} 
 &\simeq& 0.54 \left(  \frac{x_t}{x_{d (s)}}  \right) 
 \qquad \textrm{for} \quad n=0, \nonumber \\ 
 && \frac{1}{2 \ln x_t} 
    \left(  \frac{x_t}{x_{d (s)}}  \right)^2 
 \qquad \textrm{for} \quad n=1, 
\end{eqnarray}
where $x_{d (s)}$ denotes the decoupling temperature 
 in the standard cosmology. 
Similar results can be obtained for $n \geq 2$. 
Note that 
 the relic energy density in the brane world cosmology 
 can be enhanced from the one in the standard cosmology 
 if the transition temperature is low enough. 
With the help of Eqs.~(\ref{def:rho_0}) and (\ref{def:xt}), 
 we find that the enhanced energy density is proportional 
 to $M_5^{-3/2}$ (roughly $M_5^{-3}$) for $n=0$ ($n=1$). 

This is the main point discussed in Ref.~\cite{OS}. 
If the above discussion is applied 
 to detailed analysis of the relic abundance of 
 neutralino dark matter, 
 we can expect a dramatic modification 
 of the allowed region in the CMSSM.

\section{Numerical results for the neutralino relic density}

In this section, we present the results of our numerical analysis. 
We calculate the relic density of the neutralino $\Omega_{\chi}h^2$ 
in the CMSSM 
with the modified Friedmann equation Eq.~(\ref{BraneFriedmannEq}). 
For this purpose, 
we have modified the code {\sc DarkSUSY} \cite{ref:darksusy} 
so that the modified Friedmann equation is implemented. 
In evaluating the relic density, coannihilations with neutralinos, 
charginos, and sfermions are taken into account. 

The mass spectra in the CMSSM are determined by 
the following input parameters :
\begin{eqnarray}
m_0, \ \ m_{1/2}, \ \ A, \ \ \tan\beta, \ \ {\rm sgn}(\mu), 
\label{eqn:mssm-parameters}
\end{eqnarray}
where $m_0$ is the universal scalar mass, 
$m_{1/2}$ is the universal gaugino mass, 
$A$ is the universal coefficient of scalar trilinear couplings, 
$\tan\beta$ is the ratio of the vacuum expectation values 
of the two neutral Higgs fields,
and ${\rm sgn}(\mu)$ is 
the sign of the higgsino mass parameter $\mu$. 
With these input parameters, 
renormalization group equations for the CMSSM parameters 
are solved using the code {\sc Isasugra}
\cite{ref:isasugra} to obtain the mass spectra at the weak scale. 
In the present analysis, we take $A=0$ and $\mu>0$ 
with $m_t$ $=$ 178 GeV (top quark pole mass) 
and $m_b$ $=$ 4.25 GeV (bottom quark $\overline{\rm MS}$ mass at $m_b$),
and investigate 
the $M_5$ dependence of the relic density $\Omega_{\chi}h^2$ 
in the three-dimensional 
CMSSM parameter space ($m_0$, $m_{1/2}$, $\tan\beta$).  

\begin{figure}[p]
\begin{center}
\begin{minipage}{17cm}
\epsfig{file=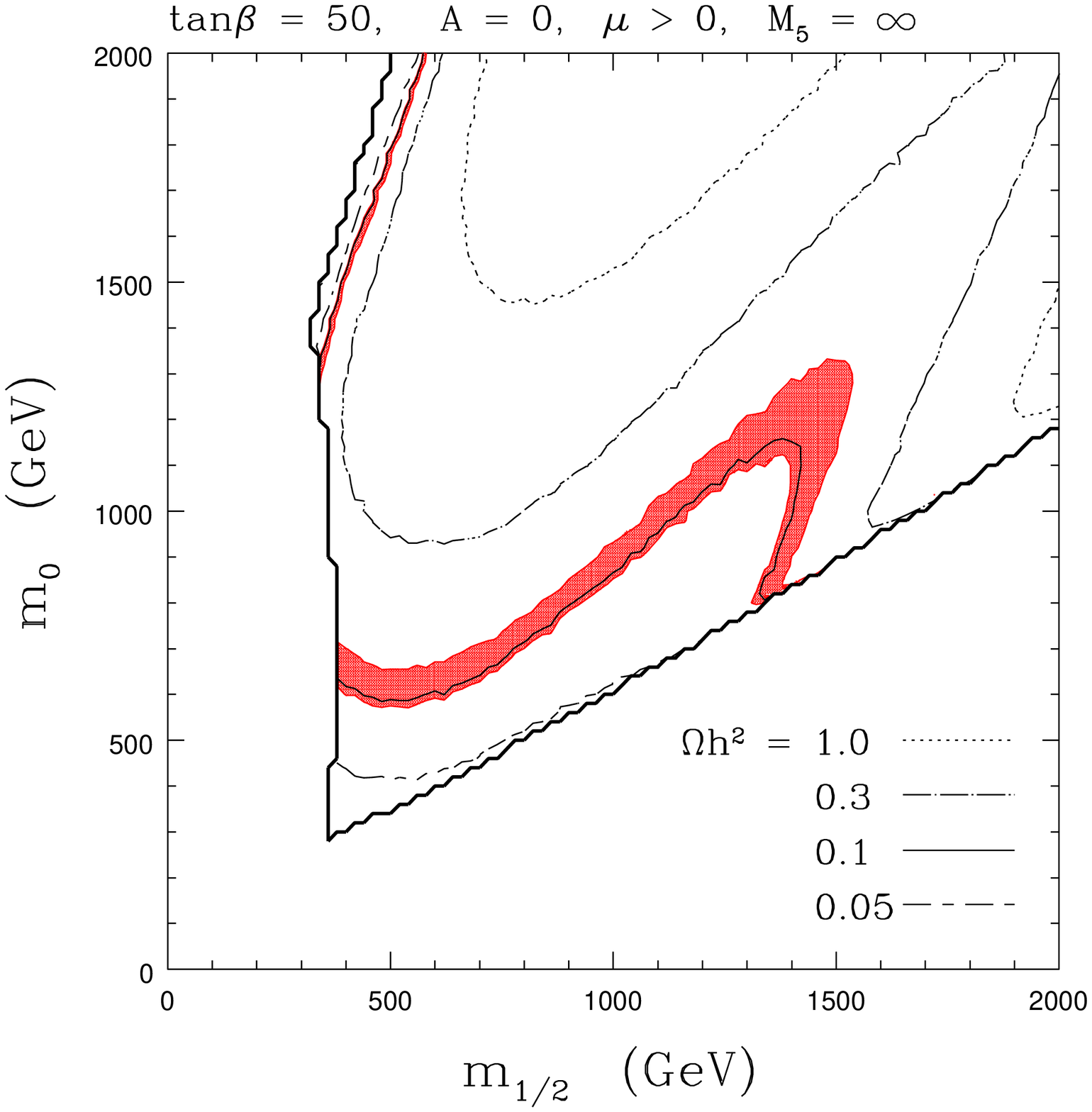,width=8cm}
\hspace*{1mm}
\epsfig{file=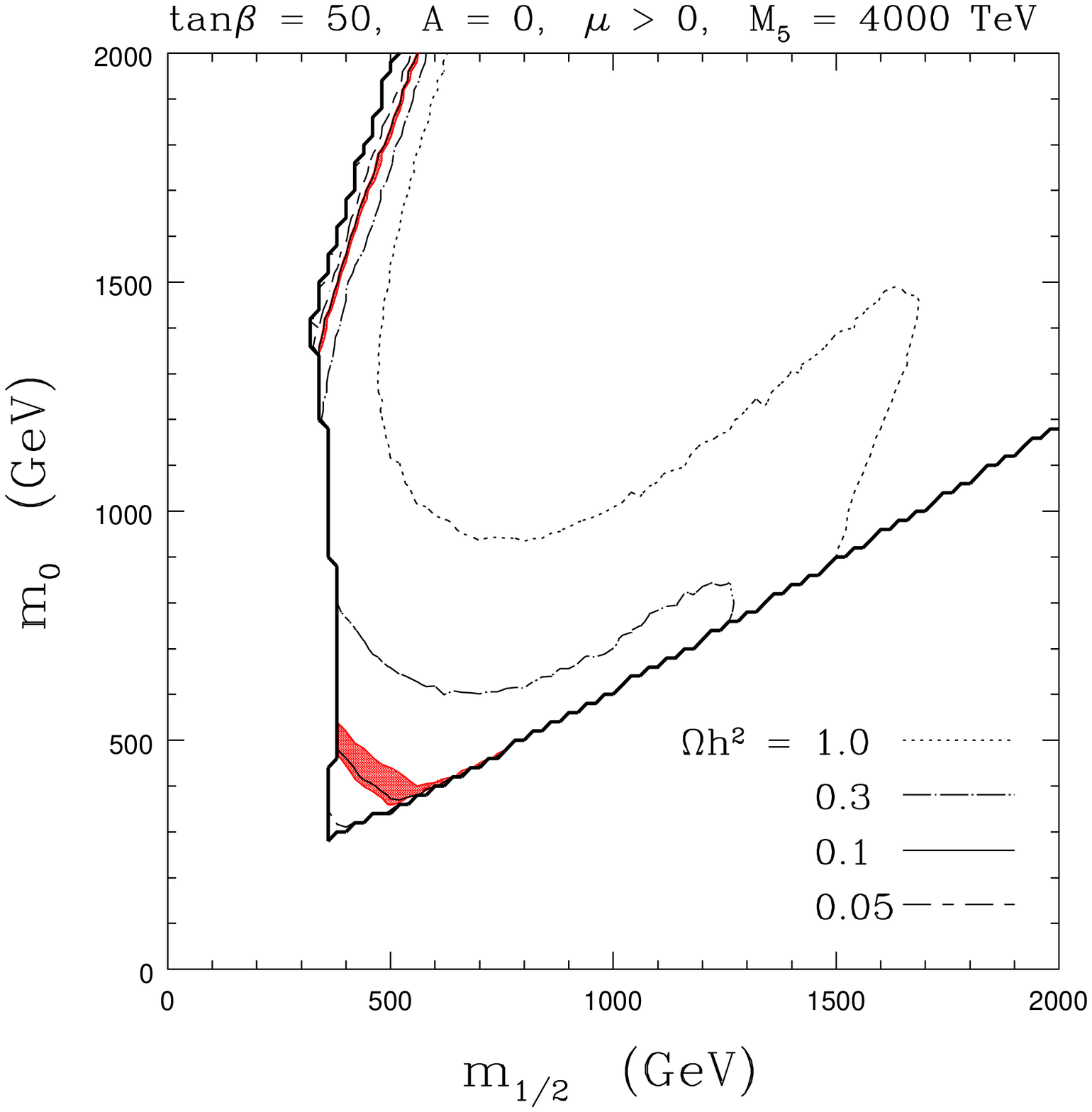,width=8cm} 
\epsfig{file=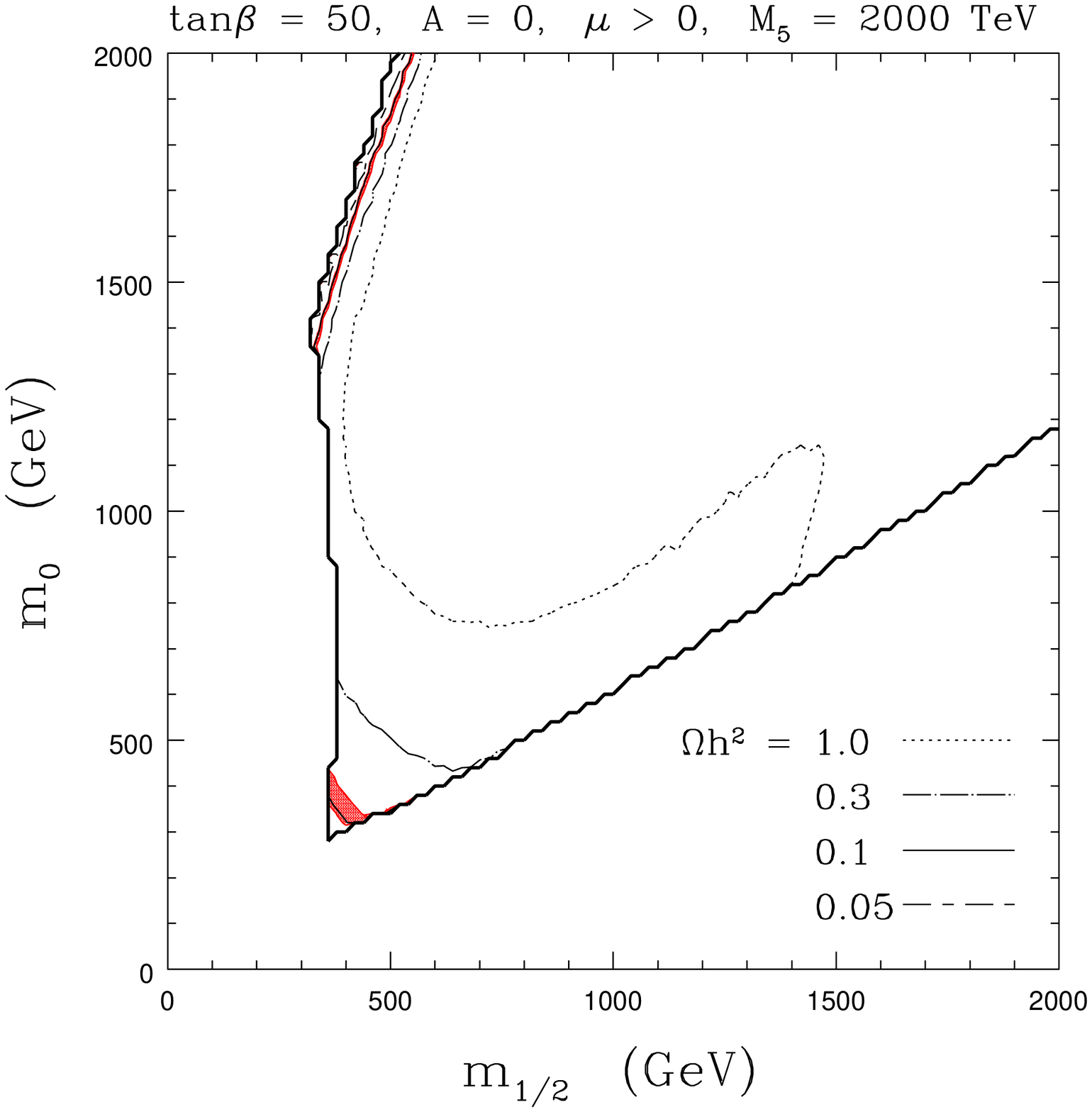,width=8cm}
\end{minipage}
\end{center}
\caption{\label{fig:tan50} 
Contours of the neutralino relic density $\Omega_{\chi} h^2$ in the 
($m_{1/2},m_0$) plane for $M_5$ $=$ $\infty$ (upper left window), 
4000 TeV (upper right window), 
and 2000 TeV (lower window)
in the case of $\tan\beta=50$, $A=0$ and $\mu>0$. 
The dotted, dashed, solid and short-dash--long-dash lines
correspond to $\Omega_{\chi} h^2$ $=$ 1.0, 0.3, 0.1 and 0.05, respectively.
The shaded regions (in red) are allowed by the WMAP constraint. 
The region outside the bold line, including the two axes, 
are excluded by experimental constraints or 
the condition for successful electroweak symmetry breaking. 
}
\end{figure}

\begin{figure}[p]
\begin{center}
\begin{minipage}{17cm}
\epsfig{file=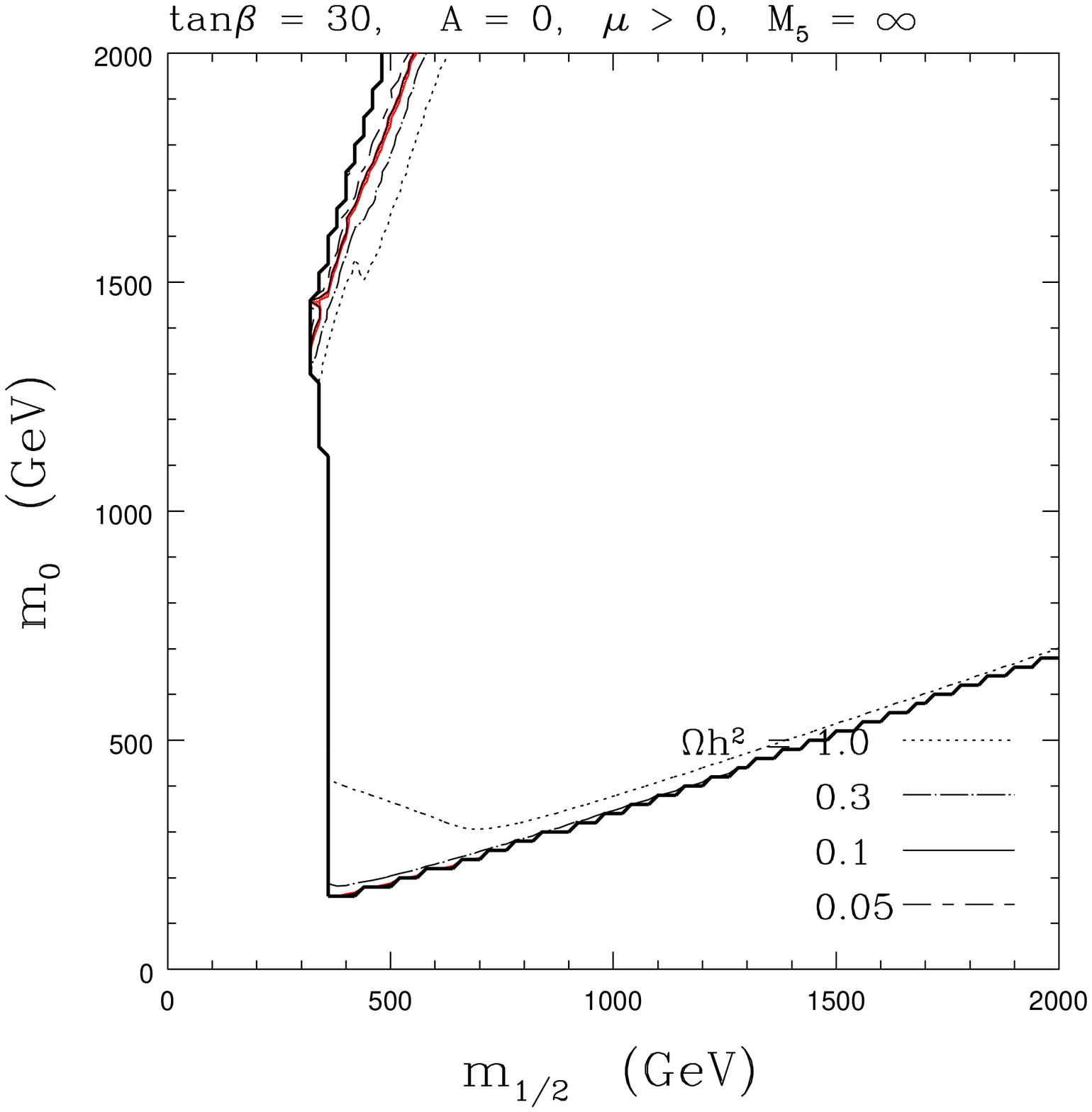,width=8cm}
\hspace*{1mm}
\epsfig{file=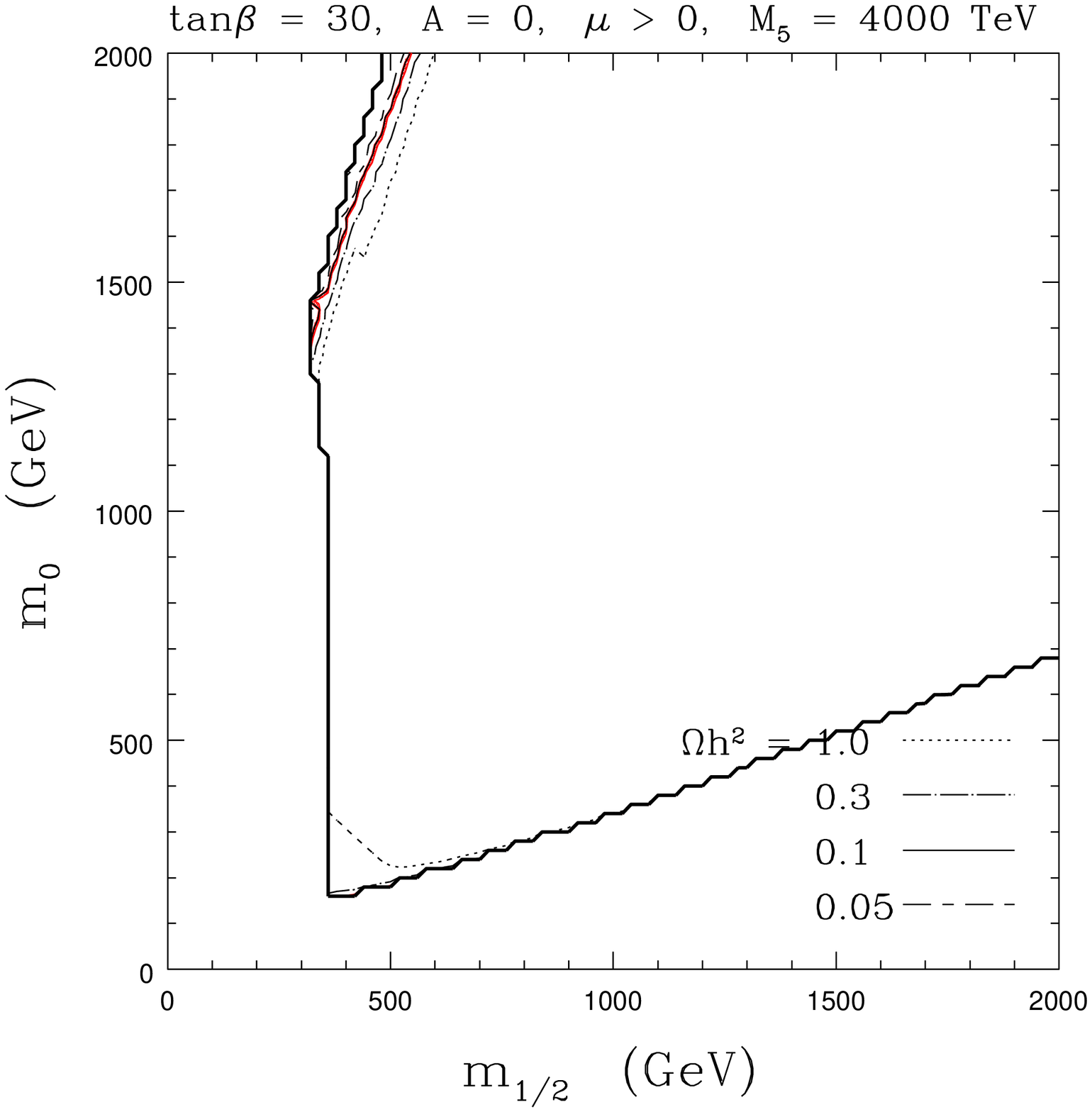,width=8cm} 
\epsfig{file=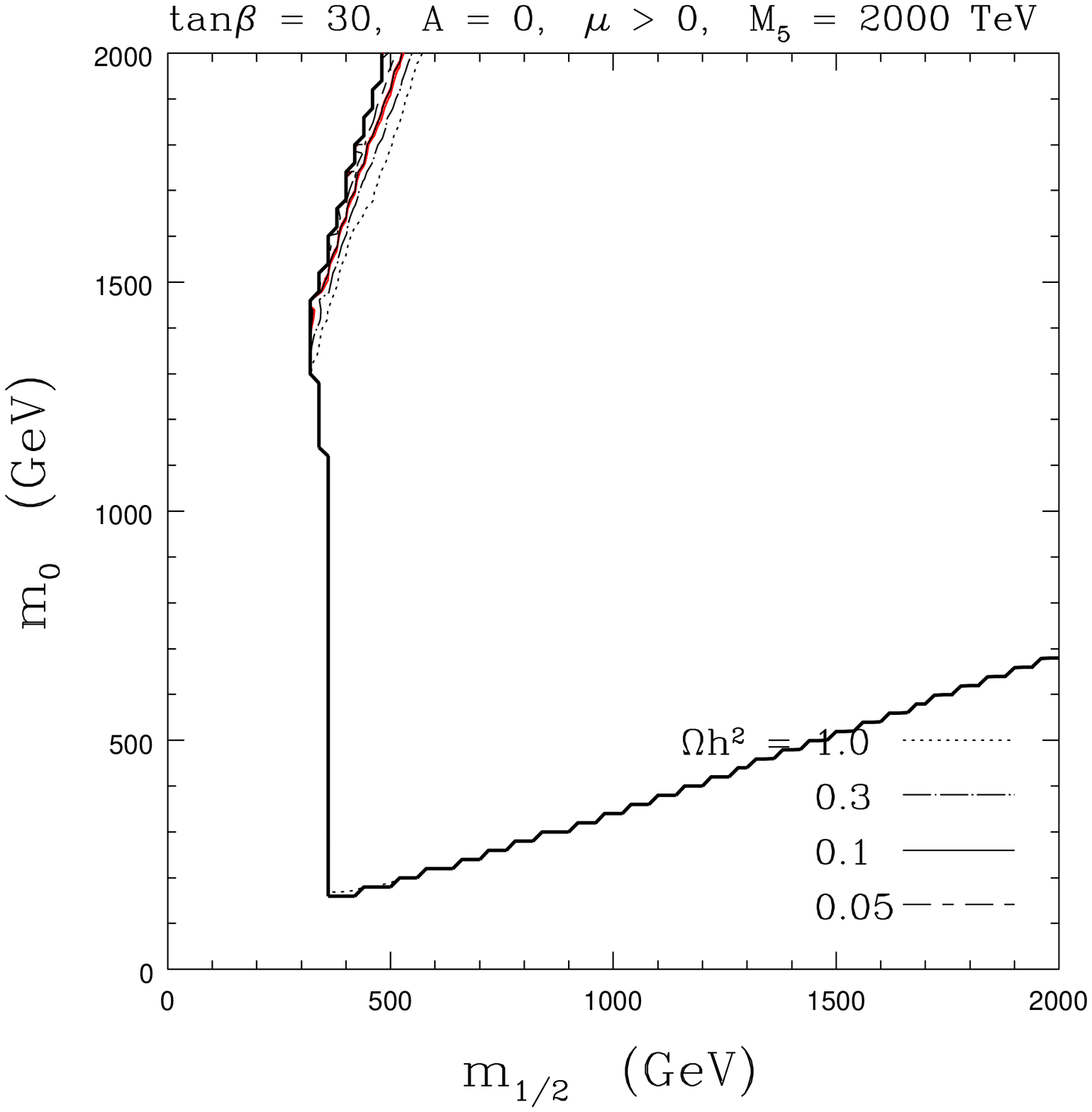,width=8cm}
\end{minipage}
\end{center}
\caption{\label{fig:tan30} 
The same as in Fig.~\ref{fig:tan50} but for $\tan\beta=30$. 
}
\end{figure}

In Fig.~\ref{fig:tan50} and Fig.~\ref{fig:tan30}, 
we show the allowed region in the ($m_{1/2},m_0$) 
plane consistent with the WMAP 2$\sigma$ allowed range 
$0.094 < \Omega_{\chi} h^2 < 0.129$. 
Figure \ref{fig:tan50} contains the contour plots of $\Omega_{\chi} h^2$
for $\tan\beta=50$, $A=0$, and $\mu>0$. 
The upper left window 
corresponds to the usual result 
in the standard cosmology ($M_5$ $=$ $\infty$). 
The dotted, dashed, solid and short-dash--long-dashed lines
correspond to $\Omega_{\chi} h^2$ $=$ 1.0, 0.3, 0.1 and 0.05, respectively.
The region among the bold line and the two coordinate axes 
 is excluded by various experimental constraints 
 (the lightest Higgs mass bound, $b\to s\gamma$ constraint, 
 the lightest chargino mass bound, etc.) \cite{CDM, PDG}
 or the condition for successful electroweak symmetry breaking (EWSB). 
In particular, 
the region $m_0 \ll m_{1/2}$ 
is excluded since the lighter stau is the LSP, 
while 
the region $m_0 \gg m_{1/2}$ 
is excluded since successful electroweak symmetry breaking 
does not occur in this region (no-EWSB region).

The shaded regions (in red) are allowed by the WMAP constraint. 
The allowed regions include 
(i) a resonance region which appears in the bulk region, 
(ii) a stau coannihilation region which appears as a narrow strip 
 along the boundary of the stau LSP region, and 
(iii) a Higgsino-like region which appears near the no-EWSB region. 
In the resonance region, 
the annihilation cross section is enhanced via heavy Higgs resonances. 
In the stau coannihilation region, 
the lighter stau is nearly degenerated with the lightest neutralino
so that the cross section is enhanced by 
coannihilation effects with the lighter stau. 
In the Higgsino-like region, 
the lightest neutralino is Higgsino-like, and 
the cross section is enhanced through coannihilation effects with 
the second-lightest neutralino and the lighter chargino, since 
$\mu$ is relatively small in this region so that 
these particles are nearly degenerated with the lightest neutralino. 

The upper right window and the lower window in Fig.~\ref{fig:tan50} 
 are the corresponding results for $M_5$ $=$ 4000 TeV  
 and 2000 TeV, respectively. 
These figures clearly indicate that, 
 as $M_5$ decreases, the allowed regions shrink significantly. 
The allowed region eventually disappears 
 as $M_5$ decreases further. 

Fig.~\ref{fig:tan30} is the similar result for $\tan\beta=30$. 
In this case, a resonance region does not exist, 
 and only narrow areas in the Higgsino-like region 
 and the stau coannihilation region 
 are allowed even in the standard case ($M_5$ $=$ $\infty$)%
\footnote{ 
 Unfortunately, the allowed coannihilation region is too narrow 
 to be clearly seen even in the upper right window. 
 See Ref.~\cite{CDM} for the figures of this region. 
 In the other two windows, there is no allowed coannihilation region. 
 }. 
As $M_5$ decreases, the allowed regions shrink significantly 
 and disappear eventually as in the case of $\tan\beta=50$. 

\begin{figure}[p]
\begin{center}
\begin{minipage}{17cm}
\epsfig{file=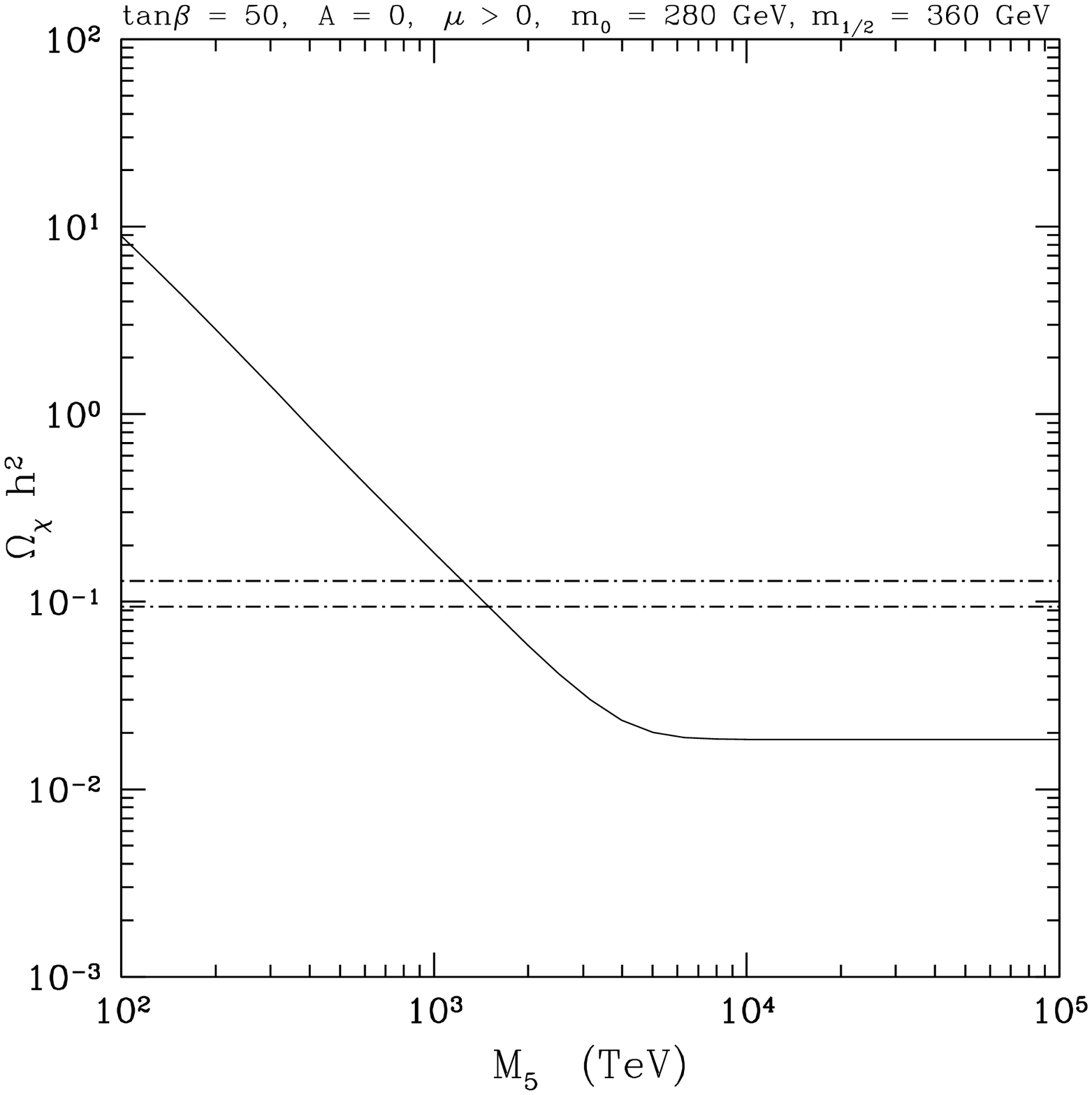,width=8cm} 
\hspace*{1mm}
\epsfig{file=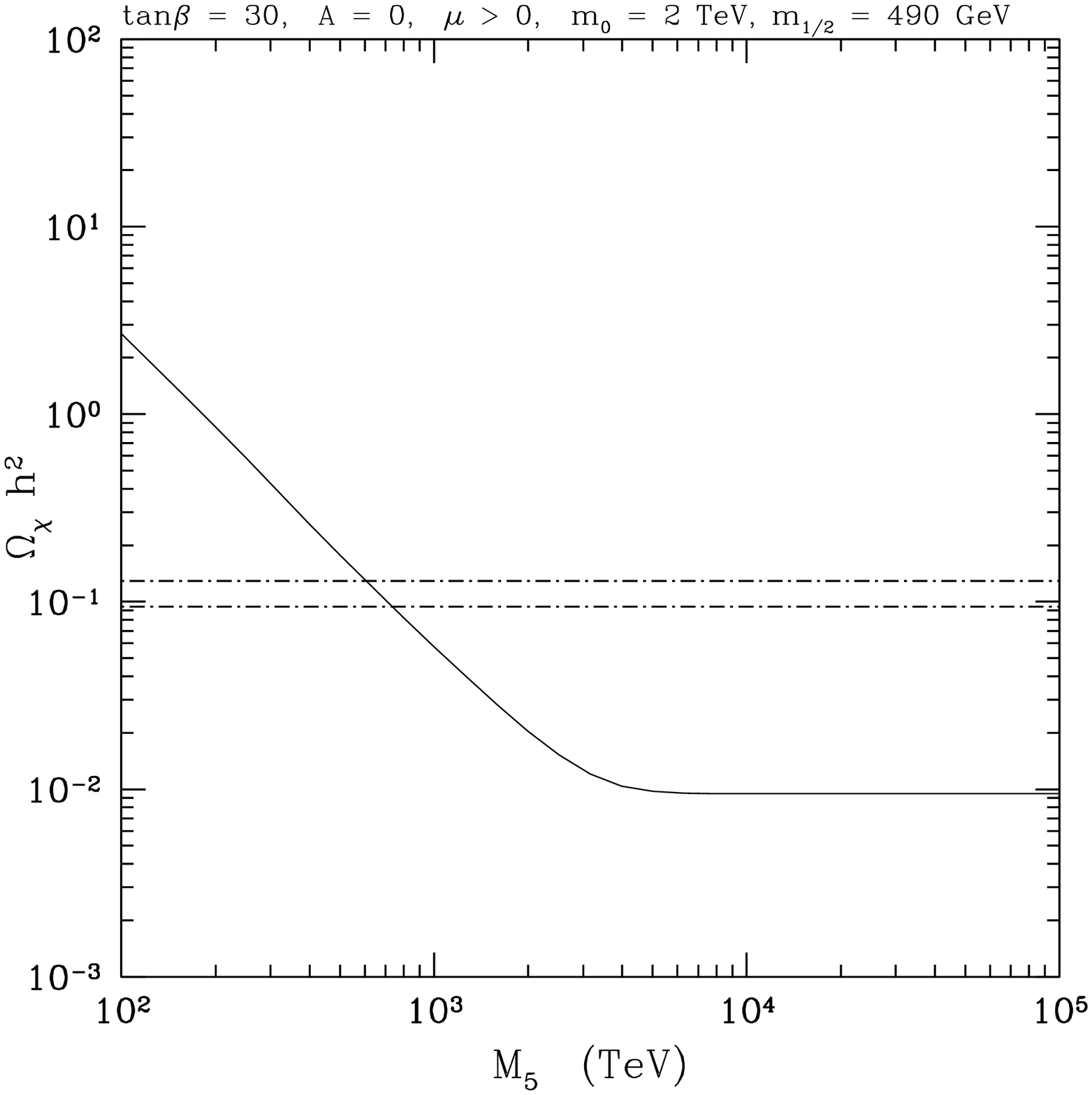,width=8cm}
\end{minipage}
\end{center}
\caption{\label{fig:m5var} 
$\Omega_{\chi} h^2$ {\it {vs.}} $M_5$ for
$\tan\beta=50$, $m_0$ $=$ 280 GeV, $m_{1/2}$ $=$ 360 GeV (left window) 
and 
$\tan\beta=30$, $m_0$ $=$ 2 TeV, $m_{1/2}$ $=$ 490 GeV (right window) 
with 
$A=0$ and $\mu>0$. 
The range of $\Omega_{\chi} h^2$ between the two dash--dotted lines 
satisfies the WMAP constraint. 
}
\end{figure}

Finally, we present sensitivity of the relic density to 
$M_5$ in Fig.~\ref{fig:m5var}, fixing $m_0$ and $m_{1/2}$ 
as well as $\tan\beta$, $A$ and sgn($\mu$). 
The left window, where we take 
 $\tan\beta=50$, $m_0$ $=$ 280 GeV and $m_{1/2}$ $=$ 360 GeV, 
 corresponds to 
 the point giving the smallest value of $\Omega_{\chi} h^2$ 
 in the small mass region ($m_0$, $m_{1/2}$ $<$ 1 TeV)
 for $\tan\beta=50$. 
The range of $\Omega_{\chi} h^2$ between the two dash--dotted lines 
satisfies the WMAP constraint. 
For large $M_5$ $\gtrsim$ $10^4$ TeV, 
the (too small) relic density ($\Omega_{\chi} h^2 \approx$ 0.02) 
in the standard case 
is reproduced independently of $M_5$. 
This is because $x_t \lesssim x_{d(s)}$ is obtained 
 for $M_5$ ($\gtrsim$ $10^4$ TeV). 
As $M_5$ decreases, however, 
the relic density $\Omega_{\chi} h^2$ starts 
to be enhanced significantly, 
and it amounts to $\approx$ 10 for $M_5$ $\approx$ 100 TeV. 
It is found that, through the enhancement,  
 an allowed region comes out for $M_5$ 
 in the range of 1000 TeV $\lesssim$ $M_5$ $\lesssim$ 1500 TeV. 

On the other hand, 
the right window, where we take 
$\tan\beta=30$, $m_0$ $=$ 2 TeV and $m_{1/2}$ $=$ 490 GeV, 
represents the point giving the smallest value of $\Omega_{\chi} h^2$ 
in the Higgsino-like region for $\tan\beta=30$. 
The $M_5$ dependence is quite similar to the case 
in the left window. 
However, 
 since the relic density $\Omega_{\chi} h^2$ $\approx$ 0.01 
 in the standard case 
 is smaller than that in the left window, 
 the WMAP allowed region is shifted 
 to smaller $M_5$ region 
 as 600 TeV $\lesssim$ $M_5$ $\lesssim$ 800 TeV. 

In Fig.~3, we can see that 
 $\Omega_{\chi}$ is proportional to about $M_5^{-2}$ 
 in the left window and $M_5^{-1.6}$ in the right window 
 for a small $M_5$ region where neutralino abundance is enhanced. 
These $M_5$ dependences of $\Omega_{\chi}$ are consistent 
 with the observation in the previous section, 
 where $\Omega_{\chi} \propto M_5^{-3/2}$ for S-wave ($n=0$)
 and $\propto M_5^{-3}$ for P-wave ($n=1$). 
This is because in the left window 
 the neutralino is Higgsino-like and 
 its pair annihilation process is dominated by S-wave ($n=0$). 
On the other hand, in the right window, 
 the neutralino is bino-like, 
 so that (for the parameters we have used in our analysis) 
 the S-wave annihilation process is somewhat suppressed 
 and the resultant $M_5$ dependence is expected to be 
 in the middle range between $M_5^{-1.5}$ and $M_5^{-3}$. 

We have produced the same contour plots 
 as Fig.~\ref{fig:tan50} and \ref{fig:tan30} 
 for $\tan\beta$ $=$ 5, 10, 20 and 40 as well. 
As in Fig.~\ref{fig:tan50} and \ref{fig:tan30}, 
 the WMAP allowed regions for each $\tan \beta $ 
 shrink and eventually disappear as $M_5$ decreases. 
It is found that, almost independently of $\tan\beta$, 
the allowed region which disappears last 
is always the Higgsino-like region. 
Then, for the Higgsino-like region, 
we have plotted the same figures as Fig.~\ref{fig:m5var}
for various $\tan\beta$ $=$ 5, 10, 20, 40 and 50. 
It turns out that the WMAP allowed range of $M_5$ 
is not so sensitive to $\tan \beta$ 
and comes out around 600 TeV $\lesssim$ $M_5$ $\lesssim$ 1000 TeV. 
Note that, since for $M_5$ smaller than this range, 
 we cannot find the WMAP allowed region, 
 $M_5 \gtrsim 600$ TeV   
 is the lower bound on $M_5$ in the brane world cosmology 
 based on the neutralino dark matter hypothesis, 
 namely, the lower bound in order for the allowed region 
 of the neutralino dark matter to exist.


\section{Conclusions}

We have studied the neutralino relic density in the CMSSM 
 in the context of the brane world cosmology.
If the five-dimensional Planck mass is low enough,
 $M_5 \lesssim 10^4$ TeV,
 the $\rho^2$ term in the modified Friedmann equation
 can be effective at the decoupling time 
 and the relic density can be enhanced. 
Therefore, we can expect that
 the allowed region in the parameter space 
 is modified from the one in the standard cosmology. 
We have presented our numerical results 
 and shown that the allowed region shrinks 
 and eventually disappears as $M_5$ decreases. 
Through the numerical analysis, 
 we have found a new lower bound on $M_5 \gtrsim 600$ TeV 
 in the brane world cosmology 
 based on the neutralino dark matter hypothesis. 
This lower bound is obtained 
 by concerning the Higgsino-like region 
 which disappears last. 
If we consider the resonance region 
or the stau coannihilation region individually 
for each $\tan \beta$ fixed, 
we can obtain a more stringent lower bound. 

%
\section*{ACKNOWLEDGMENTS}
N.O. would like to thank the Abdus Salam International Centre 
 for Theoretical Physics (ICTP), Trieste, 
 during the completion of this work. 
The works of T.N. and N.O. are supported in part 
 by the Grant-in-Aid for Scientific Research 
 (\#16740150 and  \#15740164) 
 from the Ministry of Education, Culture, Sports, 
 Science and Technology of Japan.  
O.S. is supported by the National Science Council of Taiwan 
under the grant No. NSC 92-2811-M-009-021.

%

\appendix

\section{}

In this appendix,
 we derive an approximate formula for a decoupling temperature of
 dark matter particles in brane world cosmology and
 present the effect of $\rho^2$ term on the decoupling of the dark matter.
It is shown that the decoupling temperature weakly depends on
 the transition temperature, and, thus, the resultant decoupling 
 temperature in the brane world cosmology is not so much 
 different from the one in the standard cosmology. 

Our discussion follows the strategy in \cite{Kolb}. 
We define the decoupling temperature $x_d$, in terms of $x$,
 as a temperature which satisfies $\Delta(x_d) = Y_{EQ}(x_d)$, 
 where $\Delta \equiv Y-Y_{EQ}$ is the deviation of the abundance
 from its thermal equilibrium value.
Here we examine the case where the annihilation process 
 is dominated by S-wave ($n=0$), for simplicity.
Then, the condition (in the $\rho^2$ term dominated era) 
 is expressed as 
\begin{eqnarray}
\Delta(x_d) \simeq
\frac{\sqrt{\left.\frac{\rho}{\rho_0}\right|_{T=m}}}
{2\lambda\sigma_0}
 = Y_{EQ} \simeq 0.145\frac{g}{g_{*S}}x^{3/2}e^{-x} 
\end{eqnarray}
Thus, $x_d$ is roughly written as 
\begin{eqnarray}
x_d \sim \ln\left[0.145\frac{g}{g_{*S}}2\lambda \sigma_0\frac{1}
{\sqrt{\left.\frac{\rho}{\rho_0}\right|_{T=m}}}\right]
 = x_{d(s)}-2\ln x_t  ,
\end{eqnarray}
 where
\begin{equation}
x_{d(s)} \equiv \ln\left[0.145\frac{g}{g_{*S}}2\lambda\sigma_0\right]
\end{equation}
 is the decoupling temperature in the standard cosmology \cite{Kolb}.
The second term, $-2\ln x_t$, is nothing but the effect 
 of the $\rho^2$ term, 
 and the minus sign indicates that 
 the brane world effect advances the decoupling time.

In the case that we are interested in, 
 $ x_{d(s)} < x_t < m/1\mbox{MeV}$. 
Using the neutralino decoupling temperature $ x_{d(s)} \simeq 30$, 
 normally used in the standard cosmology, 
 we obtain $7.0 < x_d < 23$ for the neutralino with mass 
$m \simeq 100$ GeV. 
The normalization used in Eq.~(\ref{critM5:100}) 
 corresponds to 
\begin{equation}
x_t \simeq \frac{g_*^{1/4}m}{\rho_0^{1/4}}
 \simeq 150\left(\frac{m}{100\, \textrm{GeV}}\right)
 \left(\frac{10^3\, \textrm{TeV}}{M_5}\right)^{3/2}. 
\end{equation}
%



\newpage

\newpage

\end{document}